\documentclass[twocolumn,           
               showpacs,            
               nopreprintnumbers,     
               aps,                 
               prd,          	    
               letterpaper,             
               groupeaddress,      
               nofootinbib,         
               tightenlines,        
               floats,floatfix      
               ]{revtex4-1}

\usepackage{graphicx}
\usepackage{dcolumn,soul}
\usepackage{bm}
\usepackage{amsmath}
\usepackage{amsfonts,amssymb}
\usepackage{color}
\usepackage{enumerate}

\begin{document}

\title{The presence of a phantom field in a Randall-Sundrum scenario}

\author{Rub\'en O. Acu\~na-C\'ardenas$^{1,2}$}
\email{roac@ifm.umich.mx}

\author{J. A. Astorga-Moreno$^3$}
\email{jesus.astorga@alumnos.udg.mx}

\author{Miguel A. Garc\'{\i}a-Aspeitia$^{4,2}$}
\email{aspeitia@fisica.uaz.edu.mx}

\author{J. C. L\'opez-Dom\'inguez$^2$}
\email{jlopez@fisica.uaz.edu.mx}

\affiliation{$^1$Instituto de F\'isica y Matem\'aticas, Universidad Michoacana de San Nicol\'as de Hidalgo. Edificio C-3, Cd. Universitaria. C. P. 58040 Morelia, Michoac\'an, M\'exico.}

\affiliation{$^2$Unidad Acad\'emica de F\'isica, Universidad Aut\'onoma de Zacatecas, Calzada Solidaridad esquina con Paseo a la Bufa S/N C.P. 98060, Zacatecas, M\'exico}

\affiliation{$^3$Centro Universitario de la Ci\'enega, Ave. Universidad 1115, C.P. 47820 Ocotl\'an, Jalisco, M\'exico}

\affiliation{$^4$Consejo Nacional de Ciencia y Tecnolog\'ia, \\ Av. Insurgentes Sur 1582. Colonia Cr\'edito Constructor, Del. Benito Ju\'arez C.P. 03940, Ciudad de M\'exico, M\'exico.}

\begin{abstract}
The presence of phantom dark energy in brane world cosmology generates important new effects, causing a premature Big Rip singularity when we increase the presence of extra dimension and  considerably competing with the other components of our Universe. The idea is based first, in only considering a field with the characteristic equation $\omega<-1$ and after that, consider the explicit form of the scalar field with a potential with a maximum (with the aim of avoid a Big Rip singularity). In both cases we study the dynamic in a robust form through the dynamical analysis theory, detailing in parameters like the deceleration $q$ and the vector field associated to the dynamical system. Results are discussed with the purpose of deeming the cosmology with a phantom field as dark energy in a Randall-Sundrum scenario.
\end{abstract}

\draft
\date{\today}
\maketitle

\section{Introduction} \label{Int}

Recent observations at high redshift with supernovae of the Type Ia \cite{riess,*perlmutter}, along with observations of anisotropies of cosmic microwave background radiation (CMB) \cite{PlanckCollaboration2013}, among others \cite{Anderson:2013zyy,*Delubac:2014aqe,*Sharov:2014voa}, show evidence of the current accelerated expansion of the Universe; suggesting the existence of a repulsive energy with the capability of accelerating the Universe, known as \emph{dark energy} (DE). These same observations also confirm that DE comprises the $\sim67\%$ of the total components and only has played role in the recent history of the Universe evolution.

In addition, from a theoretical treatment of Raychaudhuri equation, it is possible to see that in order to obtain an accelerated expansion it is necessary that the dark fluid fulfills the equation of state (EoS) $\omega<-1/3$. In this vein and with the aim to explain DE, the less \emph{expensive} candidate is the well known cosmological constant (CC), originally introduced by Einstein, but with a modern point of view regarding to its origin, expressing the EoS as $\omega=-1$ to obtain an accelerated expansion. Despite the excellent agreement of CC with observations \cite{PlanckCollaboration2013,Anderson:2013zyy,*Delubac:2014aqe,*Sharov:2014voa}, CC has a fundamental problem since we assume that it comes from the contributions of quantum vacuum fluctuations \cite{weinberg,*zeldovich,*Carroll:2000fy}, having $\sim120$ orders of magnitude of difference between the theoretical expectation value  and the observational value \cite{Carroll:2000fy}. In this sense, the theoretical community has been exploring many alternatives to control this problem, without a clear resolution so far \cite{Carroll:2000fy,weinberg,zeldovich,Martinez-Robles:2016uwa,Amendola}. However, this fundamental problem has encouraged the scientific community to propose alternative candidates for DE like the quintessence, phantom field, Chaplygin gas and extra dimensions models, among others (for a thorough review of all these alternative models see \cite{Copeland:2006wr,Amendola,Leithes}); however until now this problem remains open, with important ongoing theoretical and observational efforts with the aim of finally understand the elusive nature of DE.

As we previously mentioned, extra dimensions models are some of the most accepted candidates to understand the accelerated expansion; being a natural solution due to the straightforward way of confronting the origin of the problem. Extra dimensional models like the one proposed by Dvali, Gabadadze and Porrati (DGP) \cite{Dvali:2000hr} is one of the most promising models to solve the DE problem, because it is possible to obtain a natural threshold between four and five-dimensional physics, explaining how gravity could leak to the bulk and vice versa, imitating the actual accelerated expansion.
Other highly successful models are the Randall-Sundrum models \cite{Randall-I,*Randall-II}, 
originally created to solve the hierarchy problem between the standard model of particles (SM) and gravity. One of them is Randall-Sundrum I (RSI), which is characterized by the introduction of a five dimensional AdS compactified extra dimension between two Minkowski branes. The second one, is the Randall-Sundrum II (RSII) which has a non compactified extra dimension with the same capability of solving the hierarchy problem in a more economical way. 

Furthermore, in the cosmological context, RSII has been the most successful model due that it provides the capability of modifying the structure of Einstein's field equations. Additionally, it's important to notice that RSII leads to three new tensors: The first one is associated with the second order corrections to the energy-momentum tensor; the second one, is a tensor associated with the existence of matter in the bulk and finally a tensor that contains non-local effects associated with the Weyl's tensor \cite{sms}. In this sense, we also emphasize that the disadvantage of this model, is the necessity to introduce the DE fluid on hand, because the geometrical characteristics are not enough to obtain a natural accelerated period.

Therefore, the RSII model provides a new paradigm for the study the Universe evolution with different components. According to this idea, we propose a dynamical analysis of the modified Friedmann equations with the addition of a matter fluid (dark and baryonic) and phantom DE  in order to study the Big Rip singularity in this context. As we know from traditional literature \cite{Caldwell:2003vq}, phantom DE produces a Big Rip singularity at 22 Gyrs which can be avoided if the potential has a maximum \cite{Singh}. Another important characteristic is that phantom field minimally coupled to gravity has the sign of the kinetic term, in contrast to the ordinary scalar fields (see also \cite{Bamba:2012cp,*Nojiri:2003vn,*Nojiri:2005sx} as a complementary literature of the phantom field). Moreover, the presence of a phantom field itself in a RSII scenario will generate a more abrupt Big Rip coupled with the brane tension $\lambda$ which is the free parameter of the theory. In this sense, we establish two limits enunciated as: $\rho\gg\lambda$ which is the high energy limit (early times) and $\rho\ll\lambda$, being the low energy limit (late times). In this vein, there are several reported attempts to constraint the brane tension parameter, through Table-Top experiments \cite{Gergely}, astrophysics observations \cite{gm,*Garcia-Aspeitia:2015mwa,*Linares:2015fsa,*yo2} and cosmological analysis like BBN \cite{MaartensR} and CMB \cite{Garcia-Aspeitia:2016kak}; indeed the brane tension lies on $\lambda_{CMB}>3.44\times10^6\rm eV^4$ in the first one and $\lambda_{TT}\gtrsim138.59\times10^{48}\rm eV^4$ in the latter one\footnote{The first one is related with Cosmic Microwave Background radiation (CMB) and the latter one with Table Top experiments (TT).}, showing an enormous difference between the results, but constraining the region of possible lambda values. Setare et. al. \cite{Setare:2008mb} made a braneworld model with a non-minimally coupled panthom field where they observed that this non-minimally coupling provides a mechanism for an indirect bulk-brane gravity interaction, for late-time cosmological evolution they achieved the -1-crossing of its equation of state parameter. 

From here, we are in position to organize the paper in the following way. Sec. \ref{EM} is dedicated to construct the modified Friedmann equation from the modified Einstein's equation on the brane, as well as set the necessary condition to obtain an accelerated Universe in this theory. Following these ideas, we construct the Subsec. \ref{NA} in order to generate a numerical analysis taking into account the baryonic and the dark matter (DM) components as a dust fluid and the phantom DE, there we also discuss the possibility of an earlier Big Rip compared with GR predictions. In Sec. \ref{DS}, we revisit the dynamical system theory, in order to apply in the following sections. Sec. \ref{PDE}, is dedicated to study phantom DE through the dynamical systems theory, focusing our attention in the evolution, the vector field and the deceleration parameter, always just considering $\omega<-1$, as the main characteristic of the phantom field. In order to extend our study, we develop Sec. \ref{PF} with the aim of generate a detailed study of the phantom scalar field, using a scalar potential with a maximum; similarly, we focus our attention in the evolution, vector field and deceleration parameter in this case.
Finally, in Sec. \ref{CD}, we discuss our results and we draw important conclusions.

We will henceforth use units in which $c=\hbar=k_{B}=1$.

\section{From modified Einstein's field equation to brane cosmology} \label{EM}

We start this analysis writing the Einstein's field equation projected onto the brane as:
\begin{equation}
G_{\mu\nu}+\xi_{\mu\nu}=\kappa^2_{(4)}T_{\mu\nu} + \kappa^4_{(5)}\Pi_{\mu\nu} + \kappa^2_{(5)}F_{\mu\nu}, \label{1}
\end{equation}
where $T_{\mu\nu}$ is the four-dimensional energy-momentum tensor of the matter trapped inside the brane, $G_{\mu\nu}$ is the classical Einstein's tensor and the rest of terms on the right and left sides of the equation are explicitly given by:
\begin{subequations}
\begin{eqnarray}
&&\kappa^2_{(4)}=8\pi G_{N}=\frac{\kappa^4_{(5)}}{6}\lambda, \\
&&\Pi_{\mu\nu}=-\frac{1}{4}T_{\mu\alpha}T_{\nu}^{\alpha}+\frac{TT_{\mu\nu}}{12}+\frac{g_{\mu\nu}}{24}(3T_{\alpha\beta}T^{\alpha\beta}-T^2), \\
&&F_{\mu\nu}=\frac{2T_{AB}g_{\mu}^{A}g_{\nu}^{B}}{3}+\frac{2g_{\mu\nu}}{3}\left(T_{AB}n^An^B-\frac{^{(5)}T}{4}\right), \\
&&\xi_{\mu\nu}=^{(5)}C^E_{AFB}n_En^Fg^{A}_{\mu}g^{B}_{\nu},
\end{eqnarray}
\end{subequations}
where $\lambda$ is related with the brane tension, $\kappa_{(4)}$ and $\kappa_{(5)}$ are the four and five-dimensional coupling constants of gravity, $G_N$ is Newton's gravitational constant, $\Pi_{\mu\nu}$ represents the quadratic corrections of the energy-momentum tensor on the brane and $F_{\mu\nu}$ gives the contributions of the energy-momentum tensor in the bulk projected onto the brane through the unit normal vector $n_A$, having always in mind that latin capital letters take the values $0,1,2,3,4$. In addition $\xi_{\mu\nu}$ gives the contributions of the five-dimensional Weyl's tensor, also projected onto the brane manifold \cite{sms}.

We start the cosmological analysis proposing the traditional homogeneous and isotropic line element as:
\begin{equation}
ds^2=-dt^2+a(t)^2(dr^2+r^2(d\theta^2+\sin^2\theta d\varphi^2)),
\end{equation}
where $a(t)$ represents the scale factor and we have assumed a flat geometry, as recent observations indicate \cite{Komatsu:2011,*Planck:2015XIII,*Planck:2015XIV,*Sharov:2014voa} i.e. $k=0$. Using Eq. \eqref{1}, with matter in the brane in the form of perfect fluids and assuming no matter in the bulk, it is possible to write the modified Friedman equation and the covariant Raychaudhuri equation in the following way \cite{m2000,*Maartens:2003tw}:
\begin{subequations}
\begin{eqnarray}  \label{FLRWM}
H^2&=&\kappa^2\sum_i\rho_i\left(1+\frac{ \rho_i }{2\lambda} \right), \\
\dot{H}&=&-\frac{3\kappa^2}{2}\sum_i(\rho_i+p_i)\left(1+\frac{\rho_i}{\lambda}\right), \label{F1}
\end{eqnarray}
\end{subequations}
where $H=\dot{a}/a$ is the Hubble parameter, $\kappa^2=8\pi G_N/3=\kappa^2_{(4)}/3$ is the \emph{renamed} gravitational coupling constant and $\rho_i$ is the energy density of the different components of the Universe. Notice that $\lambda$ is the free parameter of the theory, giving the threshold between low and high energy regimes of the Universe evolution. Is it important to notice how the regime of low energy is recovered when the following ratio is applied: $\rho_i/2\lambda\to0$, recovering the traditional cosmological behavior.

In addition, the EoS necessary to accelerate the Universe satisfies the constraint
\begin{equation}
w_{p}<-\frac{1}{3}\left[\frac{1+2\rho_{p}/\lambda}{1+\rho_{p}/\lambda}\right], \label{eosacc}
\end{equation}
where in this case, the equation corresponds to a DE fluid. The previous equation can be easily calculated assuming $\ddot{a}/a>0$ in Eq. \eqref{F1}, to obtain an accelerated Universe \cite{m2000}. If we are also considering phantom DE we additionally impose the condition $\omega_p<-1$ \cite{Caldwell:2003vq}, which implies that $\rho_p/\lambda>-2$ for the phantom field.

\subsection{First Integrals} \label{NA}

First of all, we specify two fundamental quantities that are dominant in the actual stage of the Universe evolution: matter (dark and baryonic) and phantom DE, i.e. $\omega_p<-1$ \cite{Caldwell:2003vq}; as we previously mentioned, we assume that the other components are negligible for late times and we also consider non interaction between the different components i.e. non crossed terms.

Under these assumptions, the Friedmann equation can be written as:
\begin{equation}
H^2=\kappa^2\left[\frac{\rho_{0m}}{a^3}\left(1+\frac{\bar{\rho}_{0m}}{a^3}\right)+\frac{\rho_{0p}}{a^{3(1+\omega_p)}}\left(1+\frac{\bar{\rho}_{0p} }{ a^{3(1+\omega_p)}}\right) \right],
\end{equation}
where we define $\bar{\rho}_{0m}\equiv\rho_{0m}/2\lambda$, $\bar{\rho}_{0p}\equiv\rho_{0p}/2\lambda$,  which depend on the free parameter of the theory. From here, it is possible to write the equations in terms of quadratures with the aim of integrating numerically. Thus, the previous equation can be written as:
\begin{equation}
\int_{a(\tau_0)}^{a(\tau_f)}\frac{da}{\sqrt{\Omega_{0m}(a^{-1}+\bar{\rho}_{0m}a^{-4})+\Omega_{0p}(a^{7/2}+\bar{\rho}_{0p}a^5)}}=\Delta\tau, \label{intsol}
\end{equation}
where it is convenient to define the following dimensionless variables: $\Omega_{0m}\equiv\kappa^2\rho_{0m}/H_0^2$, $\Omega_{0p}\equiv\kappa^2\rho_{0p}/H_0^2$ and $\tau\equiv H_0t$.

\begin{figure}[htbp]
\centering
\begin{tabular}{cc}
\includegraphics[scale=0.3]{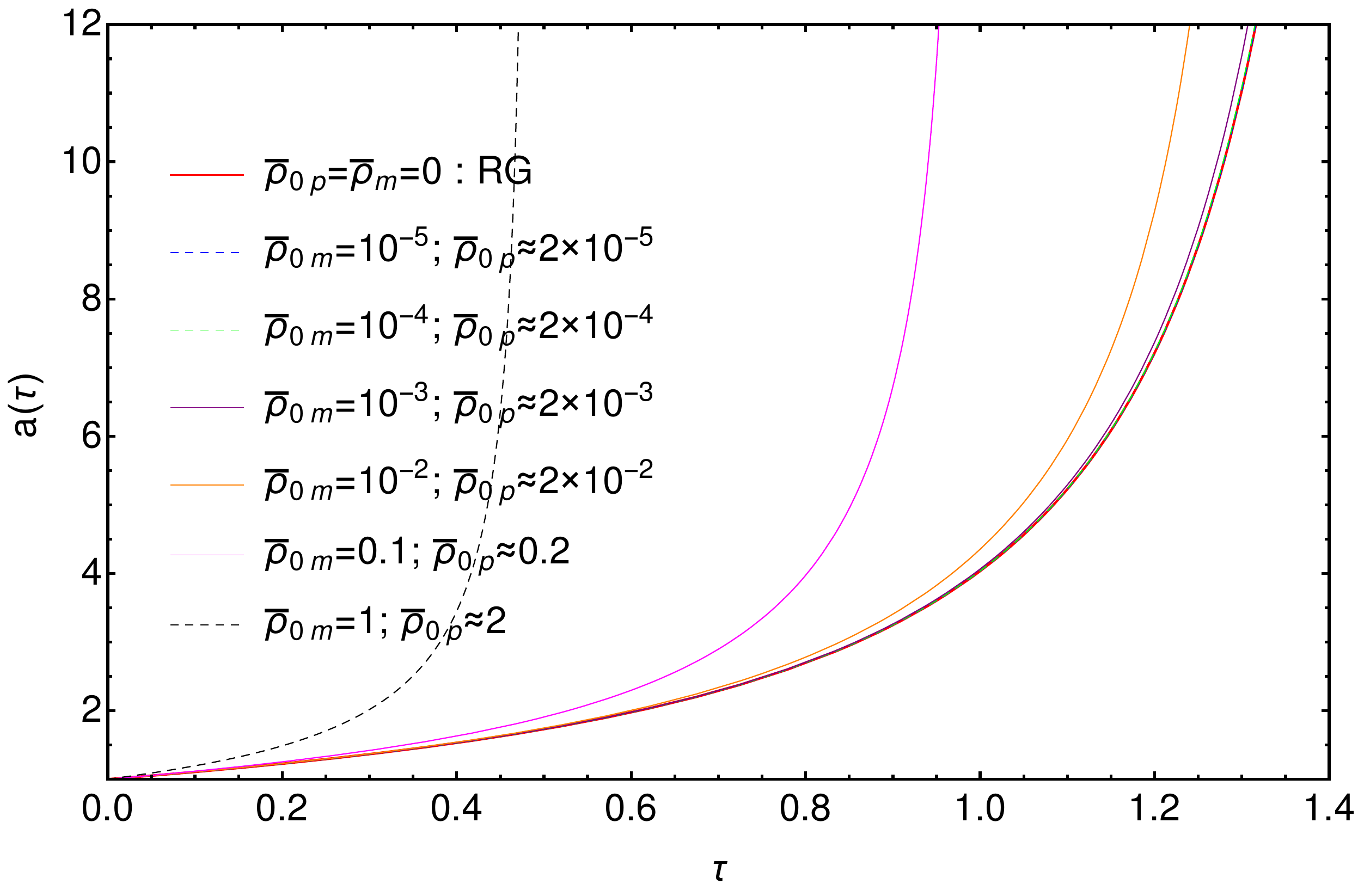} \\
\includegraphics[scale=0.3]{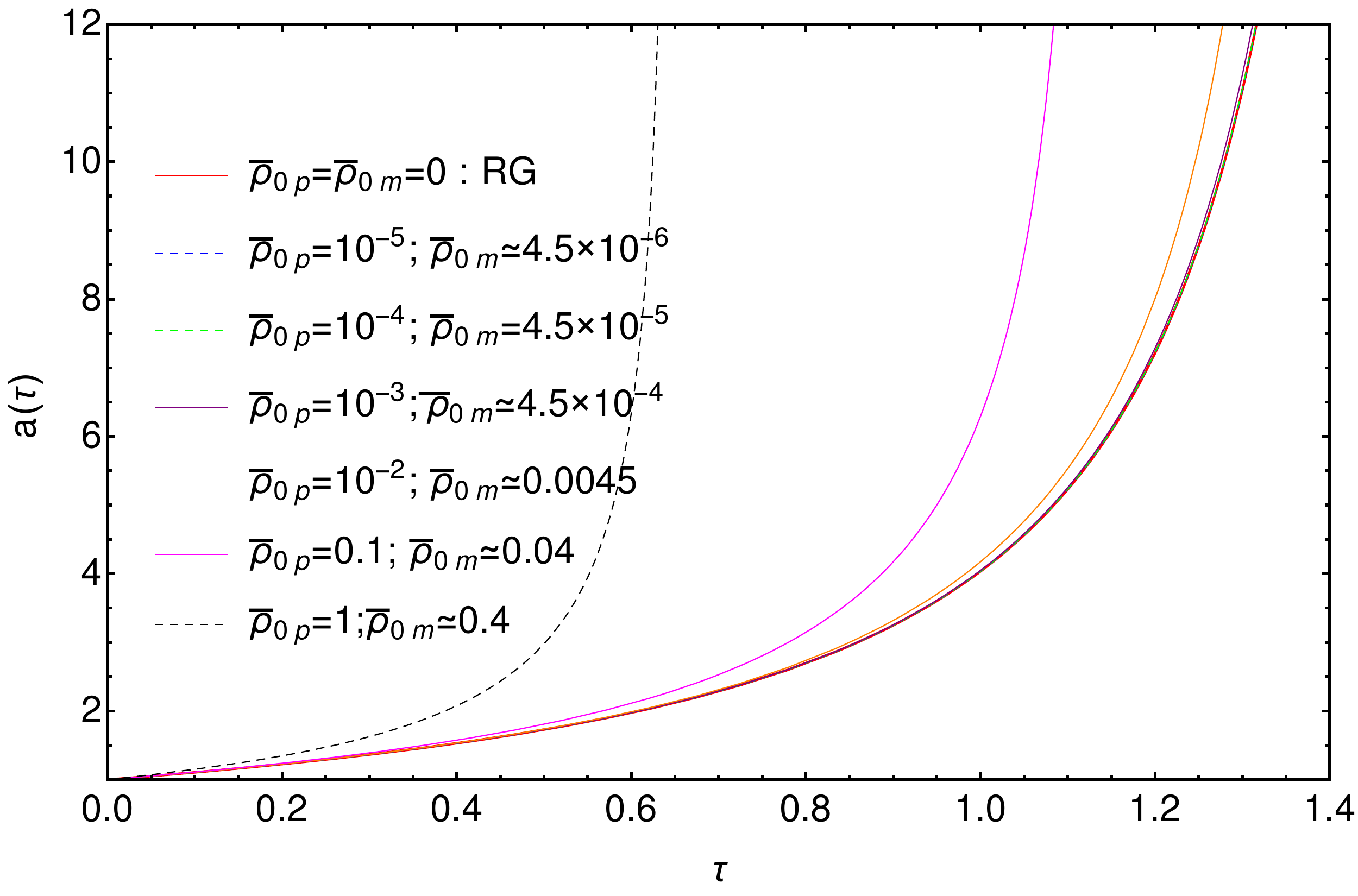}
\end{tabular}
\caption{Numerical solution of Eq. \eqref{intsol} for different values of $\bar{\rho}_{0m}$ and $\bar{\rho}_{0p}$, including the GR case with a Big Rip singularity at 22 Gyr \cite{Caldwell:2003vq}. Notice that Big Rip singularity occurs at earlier time if we increase the brane tension parameter.} 
\label{fig1}
\end{figure}

As we can see in Fig. \ref{fig1}, the Big Rip singularity occurs earlier than predicted by GR, provided that we increase the presence of extra dimensions mediated by the brane tension. In this sense, high energy in early times in the Universe evolution could have caused totally different dynamics in contrast to the expected under the presence of a phantom field.  
Moreover, the reader can verify that we reproduce the results obtained by Ref. \cite{Caldwell:2003vq} for the Big Rip singularity at 22 Gyrs, using the observational value of the Hubble constant \cite{PlanckCollaboration2013}. Thus, we notice that observations can constraint the brane tension parameter to bound the presence of extra dimensions in the case where DE is modeled by a phantom field.

\section{Revisiting a Dynamical system analysis} \label{DS}

The dynamical systems play an important role in Cosmology \cite{Copeland:2006wr} particularly in understanding the evolution of the Universe via the solutions of either the numerical or the analytical equations. Many of the tools given by this area have been widely applied in some models offering the possibility of studying different epochs of the Universe in this process (see for instance \cite{UrenaLopez:2007vz,CalderaCabral:2008bx,*Copeland:2006wr}). 

As we mention above, in the following section we present a model with phantom dark energy in a brane world to be analyzed comprehensively. However, previous to our analysis we do a theoretical  revision of dynamical systems.

First, consider the non-linear three dimensional system
\begin{subequations}
\label{bif}
\begin{eqnarray}
x^{\prime}& = & xf(x,y,z)+\alpha x, \\
y^{\prime} & = & yf(x,y,z)-\beta y,\\
z^{\prime} & = & zf(x,y,z)+\gamma z,
\end{eqnarray}
\end{subequations}
where
\begin{equation}
f(x,y,z)=-\alpha x^k+\beta y^k-\gamma z^k, \quad k \geq 1,
\end{equation}
and $\alpha,\beta,\gamma \in \mathbb{R}^{+}\setminus \{0\}$. For each selection of these parameters we have the associated critical points
\begin{equation}
s_i=(\delta_{i1},\delta_{i2},\delta_{i3}), \quad i=0,1,2,3,\label{critical}
\end{equation}
with $\delta_{ij}$ the Kronecker delta. The Jacobian matrix of the system is
\begin{widetext}
\begin{equation}
 \mathcal{J}= \left(
\begin{array}{ccc}
-\alpha(k+1) x^k+\beta y^k- \gamma z^k+\alpha& -\alpha k yx^{k-1} & -\alpha k zx^{k-1} \\
 -\beta k xy^{k-1} & -\alpha x^k+\beta(k+1) y^k- \gamma z^k-\beta & \beta k zy^{k-1}  \\
-\gamma k xz^{k-1}  & -\gamma k yz^{k-1}  & -\alpha x^k+\beta y^k- \gamma (k+1)z^k+\gamma \\
\end{array}
\right). \label{bif1}
\end{equation}
\end{widetext}
If ${\bf{x} }=(x,y,z)$ and considering a small perturbation 
\begin{equation}
{\bf{x} }\to s_i+\delta\bf{x},   \label{bif0}
\end{equation}
 we obtain the associated system $\delta\bf{x}^{\prime}=\mathcal{J}_{s_i}\delta\bf{x}$, where $\mathcal{J}_{s_i}$  is the Jacobian at the point $s_i$ and the Hartman-Grobman theorem guarantees the existence of a neighborhood for a critical point on which the flow of the system \eqref{bif} is topologically equivalent to the linearized one. Thus the eigenvalues are
\begin{eqnarray}
\lambda^{i}_1&=&\alpha-(k+1)\alpha\delta_{i1}+\beta \delta_{i2}- \gamma\delta_{i3},\label{eigen}\\
\lambda^{i}_2&=&-\beta-\alpha\delta_{i1}+(k+1)\beta\delta_{i2} -\gamma\delta_{i3} ,\label{eigen1}\\
\lambda^{i}_3&=&\gamma-\alpha\delta_{i1}+\beta \delta_{i2}- (k+1)\gamma\delta_{i3},\label{eigen2}
\end{eqnarray}
then, fixing the parameters $\beta=3/2^{k-1}$ y $\gamma=6/2^{k-1}$ we define the functions
\begin{equation}
g(i,k,\alpha)(j)=\lambda^{i}_j,    \label{bif3}
\end{equation}
where it is possible to identify the values of $\alpha$ for which $g<0$, $g>0$, $g=0$; and from the Table \ref{bif4}, we are able to identify the kind of point of the nonlinear system. 

\begin{table}[htp]
\caption{Possible values for $\alpha$ and $g$ for every $k$. }
\begin{ruledtabular}
\begin{tabular}{c c c c c c c c}
$s_{i}$ &$j$& $g<0$  &     $g>0$       & $g=0$  \\
 $i=0$                                                      &1& $(-\infty,0)$& $(0,+\infty)$ & $\alpha=0$ \\
                                                         &  2   &     $(-\infty,+\infty)$        &   &   \\
                                                         &   3   & & $(-\infty,+\infty)$  & \\
    $i=1$&1& $(0,+\infty)$& $(-\infty,0)$ & $\alpha=0$ \\
       &  2   &     $(-\infty,-\frac{3}{2^{k-1}})$        & $(-\frac{3}{2^{k-1}},+\infty)$  &  $\alpha=-\frac{3}{2^{k-1}}$ \\
        &  3  &    $(\frac{6}{2^{k-1}},+\infty)$       &  $(-\infty,\frac{6}{2^{k-1}})$   &  $\alpha=\frac{6}{2^{k-1}}$ \\
$i=2$&1& $(-\infty,-\frac{3}{2^{k-1}})$& $(-\frac{3}{2^{k-1}},+\infty)$ & $\alpha=-\frac{3}{2^{k-1}}$ \\
       &  2   &           &  $(-\infty,\infty)$   &   \\
        &  3  &           & $(-\infty,\infty)$  &   \\     
 $i=3$&1& $(-\infty,\frac{6}{2^{k-1}})$ & $(\frac{6}{2^{k-1}},+\infty)$ & $\alpha=\frac{6}{2^{k-1}}$ \\
       &  2   &      $(-\infty,\infty)$       &   &   \\
        &  3  &      $(-\infty,\infty)$       & &   \\               
\end{tabular}
\label{bif4}
\end{ruledtabular}
\end{table}
From here, we observe that for every $k$ and $\alpha \neq 0$ the critic point $s_0$ is always a saddle point, and if $\alpha$ decays at the rate $0<b < 1/2^{k-1}$, we find stability at the point $s_3$ since we have an hyperbolic system.

Now, for the system 
\begin{subequations}
\label{bif11}
\begin{eqnarray}
x^{\prime}& = & f_1(x,y,z)+xF(x,y,z), \\
y^{\prime} & = & f_2(x,y,z)+yF(x,y,z),\\
z^{\prime} & = & f_3(x,y,z)+zF(x,y,z),
\end{eqnarray}
\end{subequations}
with $f_1,f_2,f_3,P,Q \in  \mathbb{R}[x,y,z]$ (the ring of polynomials in three variables with real coefficients) that satisfies $f_1(\mathbf{0})=f_2(\mathbf{0})=f_3(\mathbf{0})=P(\mathbf{0})=0$ and $F=P/Q$ an element of the set of rational function in three variables $\mathbb{R}(x,y,z)$, with the possibility that it is not defined at the origin, also $f_1, f_2, f_3, P$ are not irreducible polynomials. Taking in mind the nonempty set
\begin{equation}
\mathcal{Z}(\mathfrak{a})=\{\mathbf{x}\in  \mathbb{R}^{3} :  f_1(\mathbf{x})=f_2(\mathbf{x})=f_3(\mathbf{x})=P(\mathbf{x})=0\},
\end{equation}
with $\mathfrak{a}=\left(f_1,f_2,f_3,P\right) \mathbb{R}[x,y,z]$ being an ideal associated to the system. For the ideal
\begin{equation}
 \mathcal{I}(\mathcal{Z}(\mathfrak{a}))=\{p \in  \mathbb{R}[x,y,z] :  p(\mathbf{x})=0, \forall \mathbf{x} \in \mathcal{Z}(\mathfrak{a})\},
\end{equation}
if $f^{n}\in\mathcal{I}(\mathcal{Z}(\mathfrak{a}))$ for some $n \in  \mathbb{N}$ then, $f \in  \mathcal{I}(\mathcal{Z}(\mathfrak{a}))$, because $\mathbb{R}[x,y,z]$ is an integer domain, showing the inclusion $\sqrt{\mathfrak{a}} \subset  \mathcal{I}(\mathcal{Z}(\mathfrak{a}))$, where $\sqrt{\mathfrak{a}}$ denotes the radical of $\mathfrak{a}$. 
Now, let $f \in  \mathcal{I}(\mathcal{Z}(\mathfrak{a}))\setminus\{0\}$ and consider the ideal 
\begin{equation}
\mathfrak{b}=\left(f_1,f_2,f_3,P,(wf-1)\right) \mathbb{R}[x,y,z,w],
\end{equation}
in the ring $\mathbb{R}[x,y,z,w]$. We notice that $\mathcal{Z}(\mathfrak{b})=\emptyset$, hence $\mathfrak{b}=\mathbb{R}[x,y,z,w]$ so, there are polynomials $p_1,p_2,p_3,p_4,p_5$ such that
\begin{equation}
1=p_1f_1+p_2f_2+p_3f_3+p_4P+p_5(wf-1).
\end{equation}
Now, considering the ring $\mathbb{R}(x,y,z)[w]$ with $w=1/f$, we have
\begin{eqnarray}
1=p_1\left(\mathbf{x},\frac{1}{f}\right)f_1+p_2\left(\mathbf{x},\frac{1}{f}\right)f_2+\nonumber\\
p_3\left(\mathbf{x},\frac{1}{f}\right)f_3+p_4\left(\mathbf{x},\frac{1}{f}\right)P,
\end{eqnarray}
and for some $k \in  \mathbb{N}$
\begin{equation}
f^{k}=q_1f_1+q_2f_2+q_3f_3+q_4P,
\end{equation}
where $q_1,q_2,q_3,q_4 \in\mathbb{R}[x,y,z]$, obtaining the inclusion $ \mathcal{I}(\mathcal{Z}(\mathfrak{a}))  \subset  \sqrt{\mathfrak{a}} $ and the equality 
\begin{equation}
 \mathcal{I}(\mathcal{Z}(\mathfrak{a}))=\sqrt{\mathfrak{a}}.  
\end{equation}

Using this result, a set of critical points for \eqref{bif11} is $\mathcal{Z}(\mathfrak{a})\setminus\{0\}$, being a possibility the points in Eq. (\ref{critical}) as
\begin{enumerate}[(a)]
\item{$s_1$  if 
\begin{equation}
f_1(x,0,0)Q(x,0,0)+xP(x,0,0)\sim_{a} f_i(x,0,0)\quad i=2,3.\nonumber
\end{equation}
}
\item{$s_2$ if 
\begin{equation}
f_2(0,y,0)Q(0,y,0)+yP(0,y,0)\sim_{b} f_i(0,y,0) \quad i=1,3.\nonumber
\end{equation}
}
\item{$s_3$ if 
\begin{equation}
f_3(0,0,z)Q(0,0,z)+zP(0,0,z)\sim_{c}f_i(0,0,z) \quad i=1,2,\nonumber
\end{equation}
}
\end{enumerate}
with $a,b,c \neq 0$ and $\sim_{r}$ denotes the equivalence relation for polynomial functions in one variable that vanish at the point $r$.  

\section{Phantom Dark Energy in a RS scenario} \label{PDE}

We now start rearranging the Eq. \eqref{FLRWM} in the form:
\begin{equation}
H^{2}=\frac{8\pi G_N}{3}\sum_{i}\left(\rho_{i}+\frac{\rho_{i}^{2}}{2\lambda}\right),\label{eq:EF_01}
\end{equation}
where, redefining the expression $\bar{\rho}_{i}\equiv\rho_{i}^{2}/2\lambda$, it is possible to write:
\begin{equation}
H^{2}=\frac{8\pi G_N}{3}\sum_{i}\left(\rho_{i}+\bar{\rho}_{i}\right). \label{eq:EF_02}
\end{equation}
In order to visualize the dynamic equations we propose two different methods. Both methods will generate the same dynamical information but in some cases we will choose one over the other, due to the different information we will be able to obtain from each of them. 

\begin{enumerate}[(a)]
\item Method 1. When the dimensionless variables are constrained into a four dimensional sphere, with Friedman constraint $1=\sum_{i}\left(x_i^2+y_i^2\right)$ and dimensionless variables:
\begin{equation}
\begin{gathered}x_{i}^2\equiv\frac{8\pi G_N}{3H^{2}}\rho_{i}, \;\;\;
y_{i}^2\equiv\frac{8\pi G_N}{3H^{2}}\bar{\rho}_{i}.
\end{gathered}
\label{eq:xi-yi_01}
\end{equation}
Applying the Friedmann constraint, the dynamical equations reduces to:
\begin{subequations}
\label{SDC1}
\begin{eqnarray}
\frac{2x_{2}^{\prime}}{3x_2} & = & -\frac{3}{2}x_2^2+y_1^2-2y_2^2+\frac{3}{2},\\
\frac{2y_{1}^{\prime}}{3y_1} & = & -\frac{3}{2}x_2^2+y_1^2-2y_2^2-1,\\
\frac{2y_{2}^{\prime}}{3y_2} & = & -\frac{3}{2}x_2^2+y_1^2-2y_2^2+2.
\end{eqnarray}
\end{subequations}

We notice that this is the dynamical system \eqref{bif} with $k=2$ and $\alpha=9/4$.

\item Method 2. When the dimensionless variables are constrained into a four dimensional plane, with Friedman constraint $1=\sum_{i}\left(x_i+y_i\right)$ and dimensionless variables:

\begin{equation}
\begin{gathered}x_{i}\equiv\frac{8\pi G_N}{3H^{2}}\rho_{i}, \;\;\;
y_{i}\equiv\frac{8\pi G_N}{3H^{2}}\bar{\rho}_{i}.
\end{gathered}
\label{eq:xi-yi_01}
\end{equation}
\end{enumerate}
In the same way as in method 1, the Friedmann constraint helps us to reduce the dynamical equations, as in the the case $k=1$ in \eqref{bif} with $\alpha=9/2$:

\begin{subequations}
\label{SDC}
\begin{eqnarray}
\frac{x_{2}^{\prime}}{3x_2} & = & -\frac{3}{2}x_{2}+y_{1}-2y_{2}+\frac{3}{2},\\
\frac{y_{1}^{\prime}}{3y_1} & = & -\frac{3}{2}x_{2}+y_{1}-2y_{2}-1,\\
\frac{y_{2}^{\prime}}{3y_2} & = & -\frac{3}{2}x_{2}+y_{1}-2y_{2}+2.
\end{eqnarray}
\end{subequations}
In both cases, the primes denote an e-folding derivative $N=\ln(a)$, where we also made use of the fact that $\omega_{m,(DM)}=\omega_{1}=0$ and $\omega_p=\omega_{2}=-3/2$, due that we had assumed that the only components of the Universe are phantom DE and matter (baryonic and DM). Notice that the choice of the phantom EoS is based on Planck satellite observations \cite{Planck:2015XIII,*Planck:2015XIV,*Sharov:2014voa}.

\begin{figure*}[h]
\centering
\begin{tabular}{cc}
\includegraphics[scale=0.35]{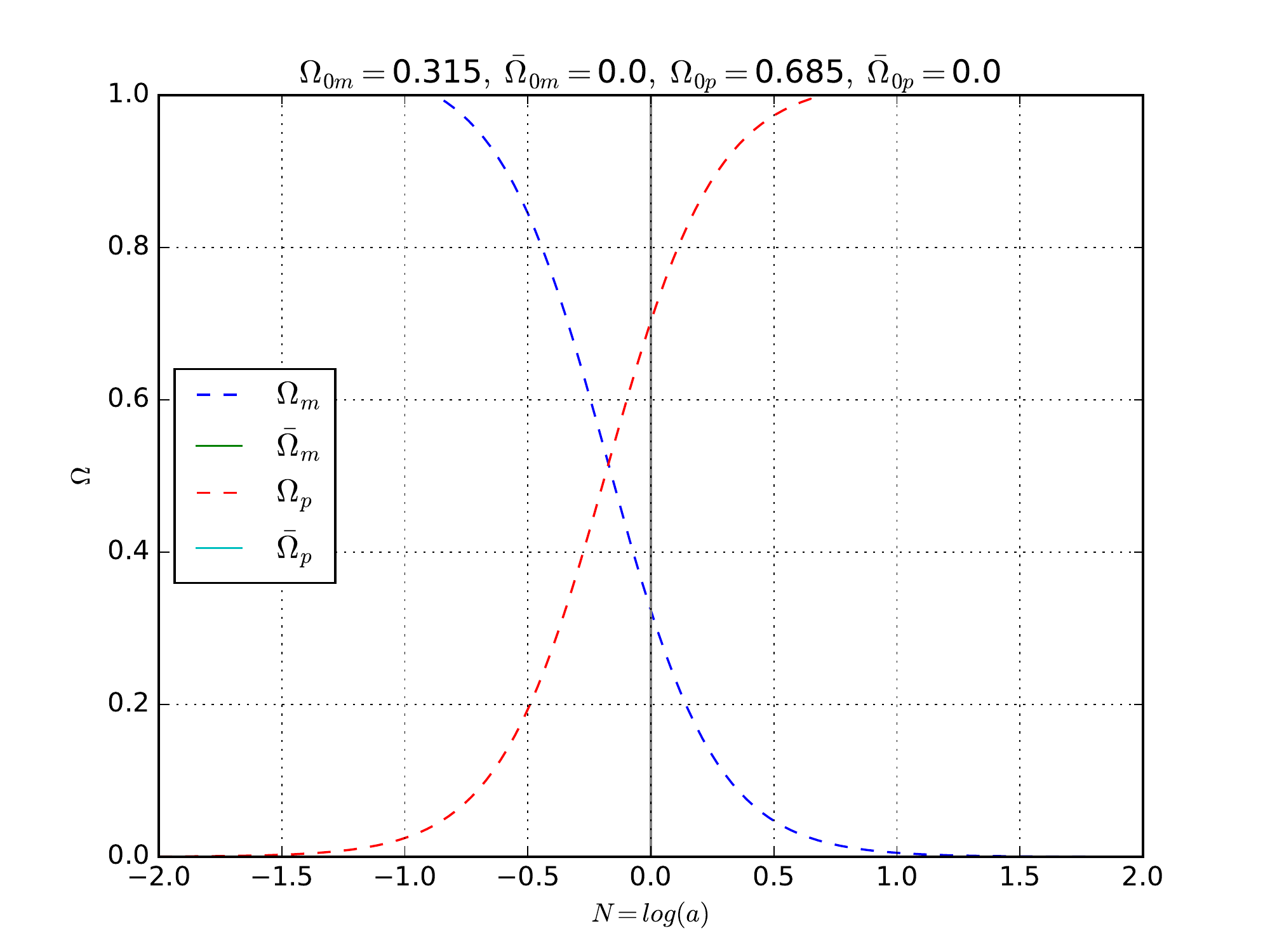} 
\includegraphics[scale=0.35]{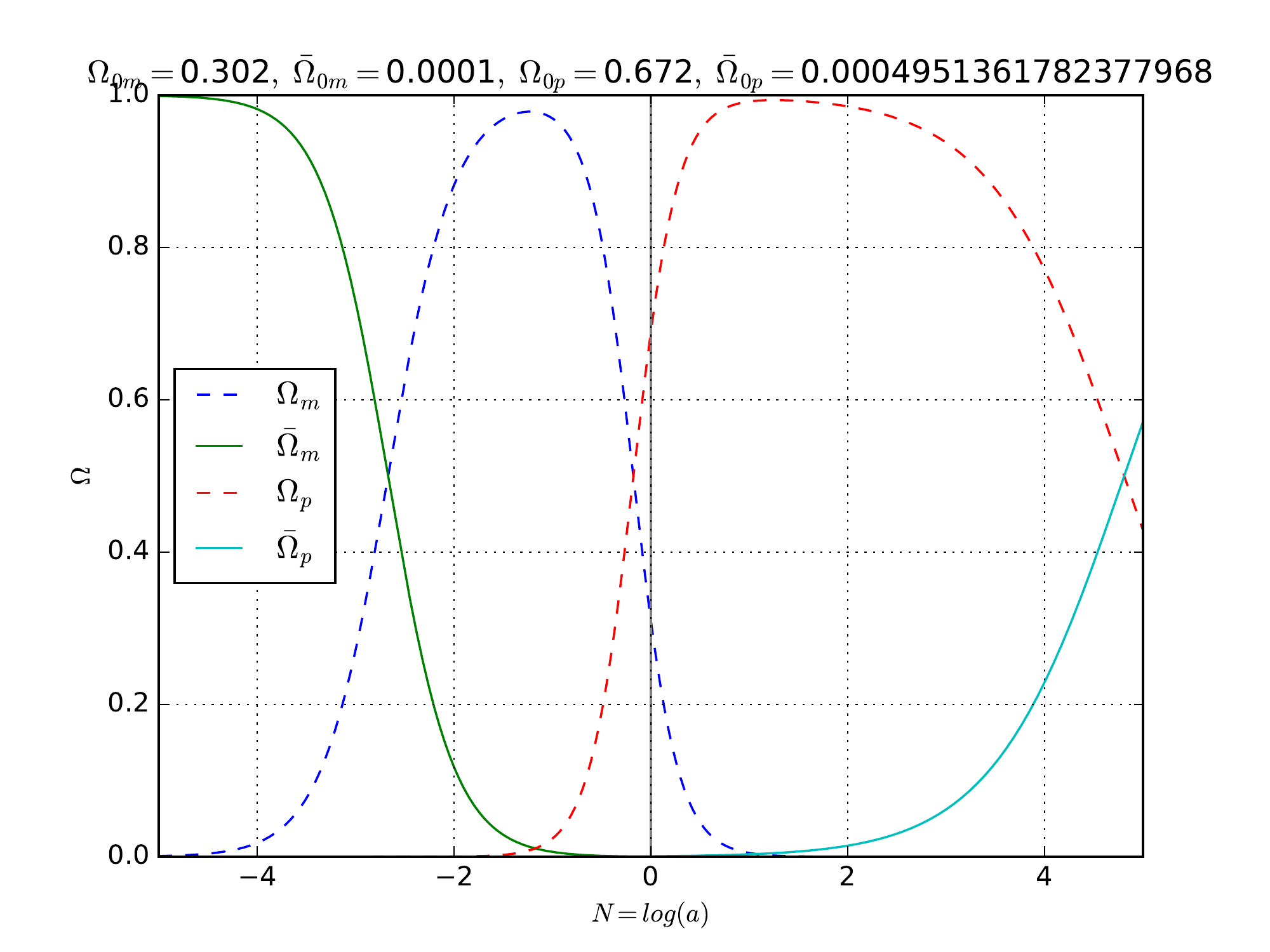} \\
\includegraphics[scale=0.35]{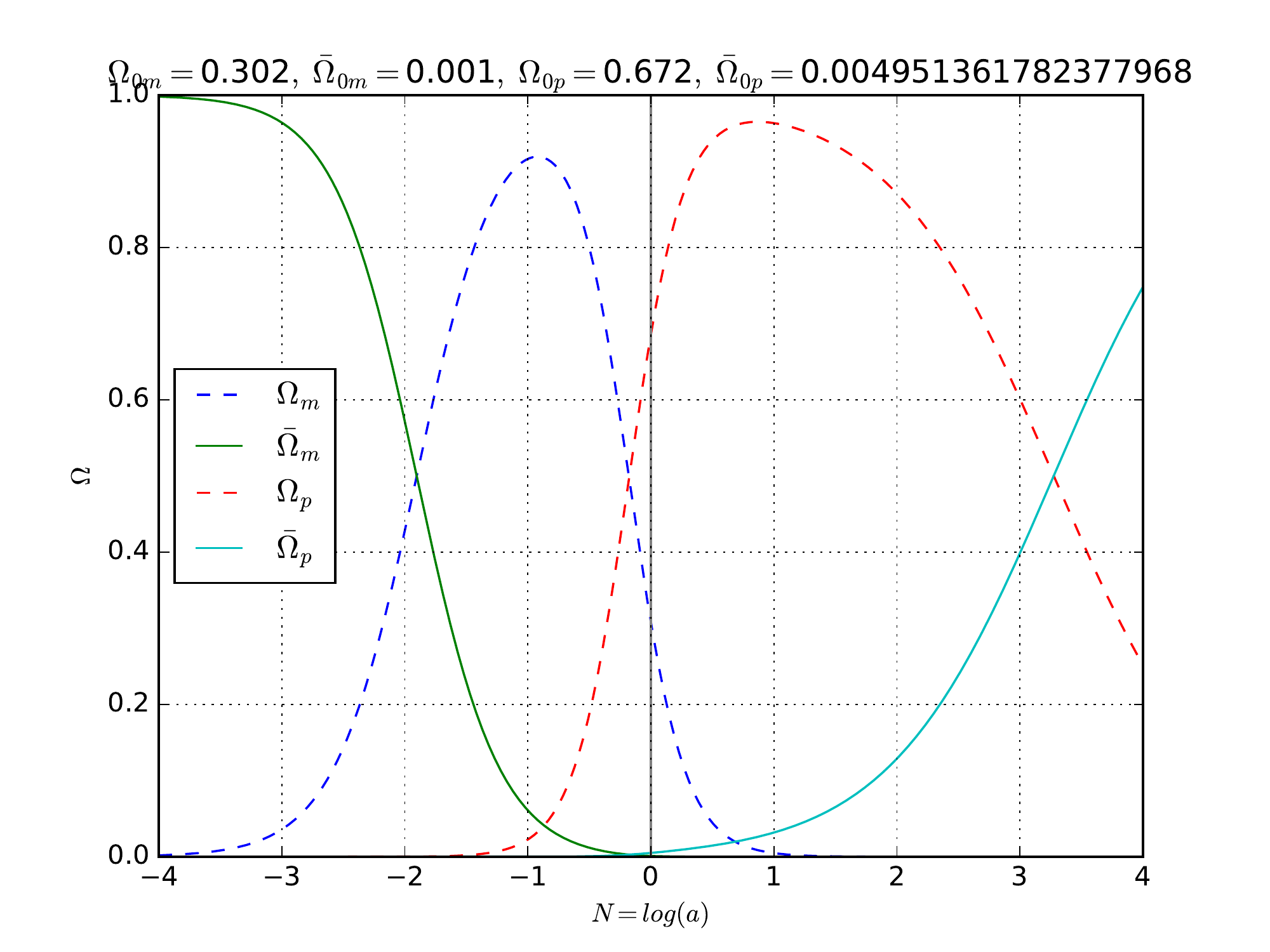} 
\includegraphics[scale=0.35]{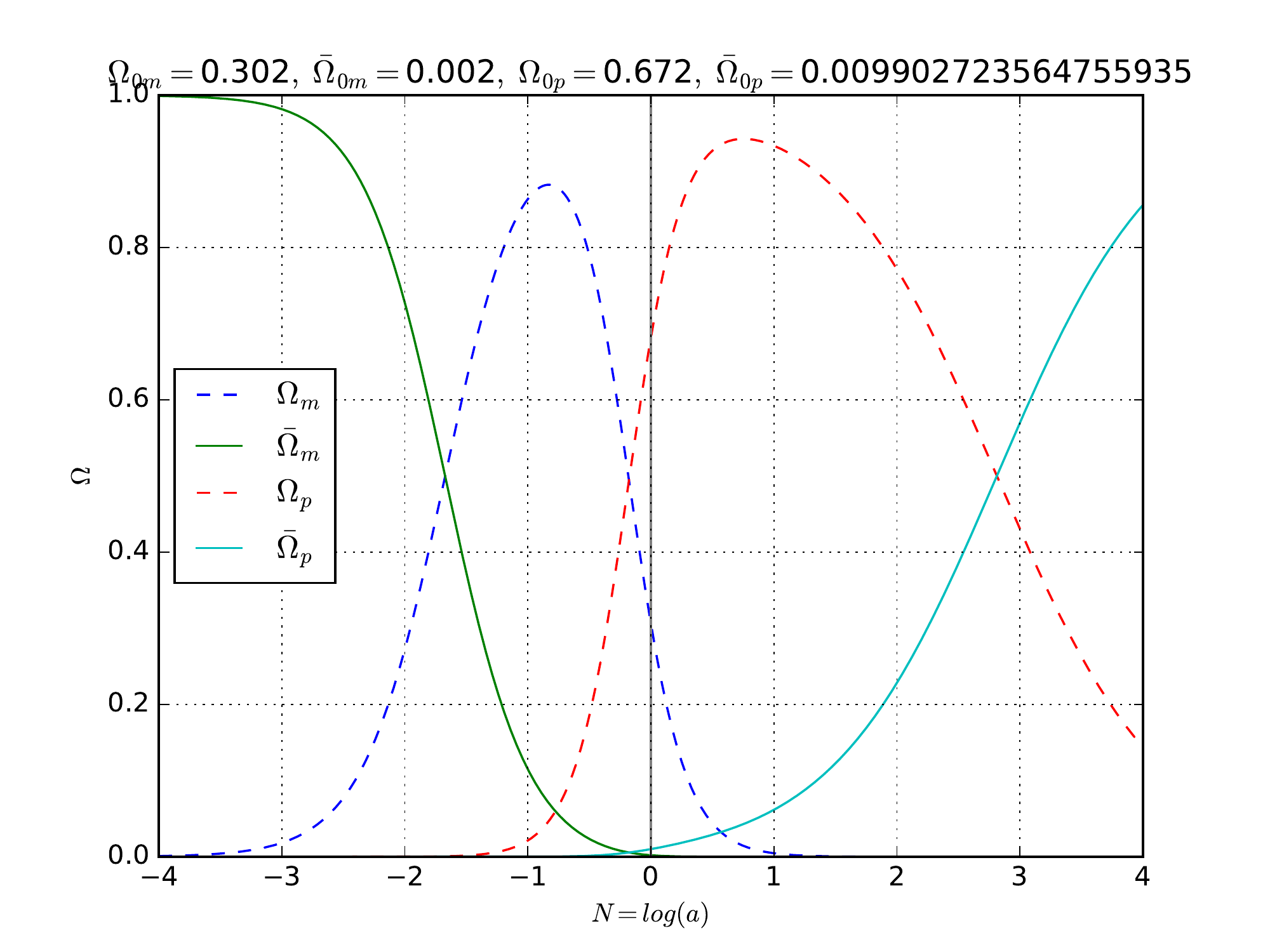} \\
\includegraphics[scale=0.35]{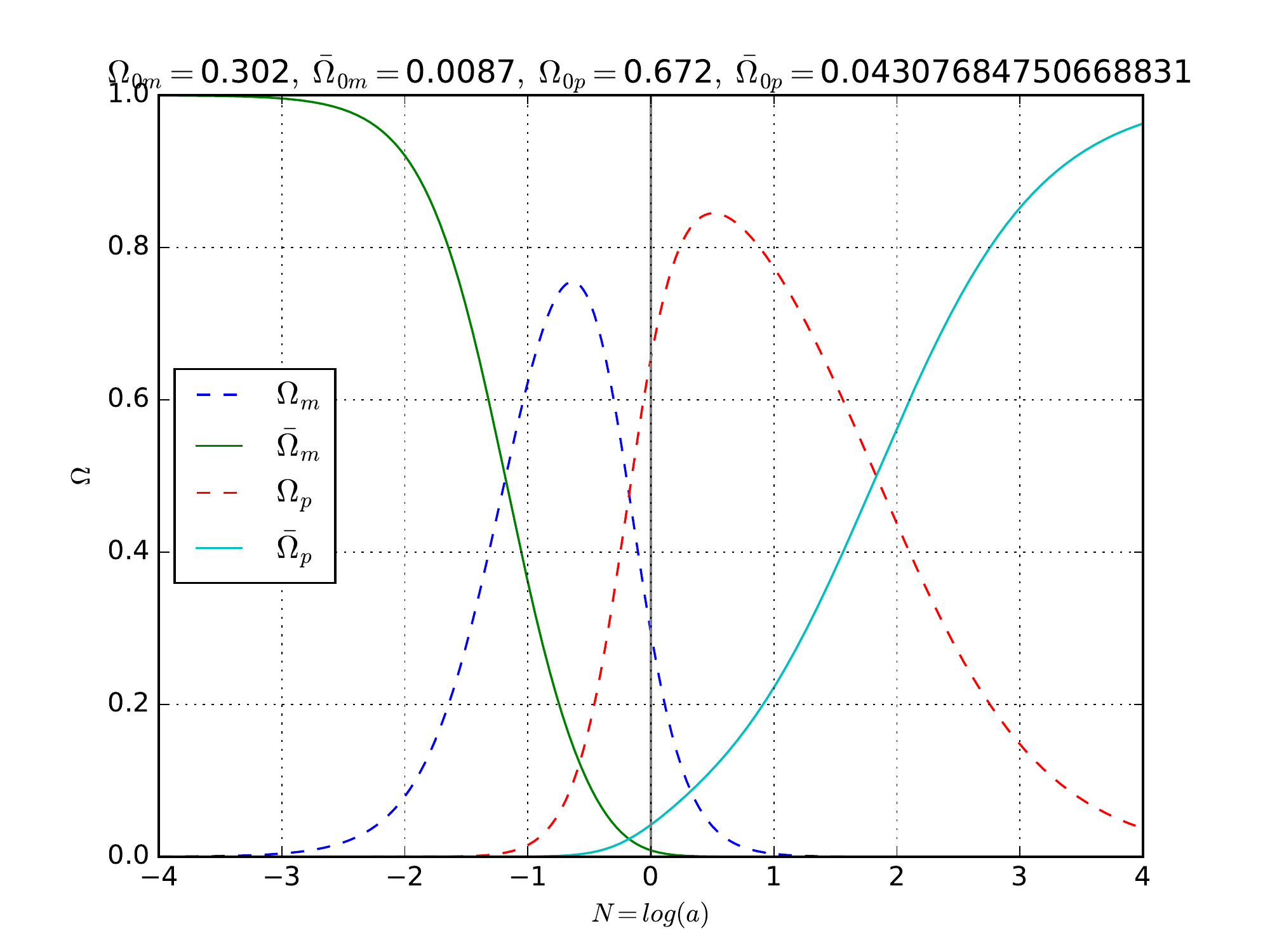}
\end{tabular}
\caption{Dynamical analysis of Eqs. \eqref{SDC1} with appropriate initial conditions for $\Omega_{0m}$, $\Omega_{0p}$, $\bar{\Omega}_{0m}$ and $\bar{\Omega}_{0p}$; the last two equations are subject to the constraints provided by the Planck satellite \cite{PlanckCollaboration2013}. In all the cases there is a notorious domination of the phantom field coupled with branes at later times while matter coupled with branes dominates the earlier stages of the Universe evolution, as can be expected.} 
\label{fig2}
\end{figure*}

For instance, the dynamical system represented by Eqs. \eqref{SDC1} can be solved numerically, which graphical solutions are shown in Figs. \ref{fig2}, establishing the initial conditions for $\Omega_{0m}$ and $\Omega_{0p}$ through the Planck satellite constraints \cite{PlanckCollaboration2013}, while the other initial conditions for $\bar{\Omega}_{0m}$ and $\bar{\Omega}_{0p}$, can be also constrained with \cite{PlanckCollaboration2013} having a permitted region to manipulate the density parameters coupled by the brane tension. Here we separate the different components with the aim of visualizing the behavior. As the reader can observe, the phantom DE coupled with branes dominates in later stages of Universe evolution while in similar conditions matter coupled with branes dominates in earlier stages of Universe; in this sense, it is important to give a more restrictive constriction on the brane tension parameter based on observations, in order to elucidate the effects of extra dimensions.

\begin{figure}[htbp]
\centering
\begin{tabular}{cc}
\includegraphics[scale=0.5]{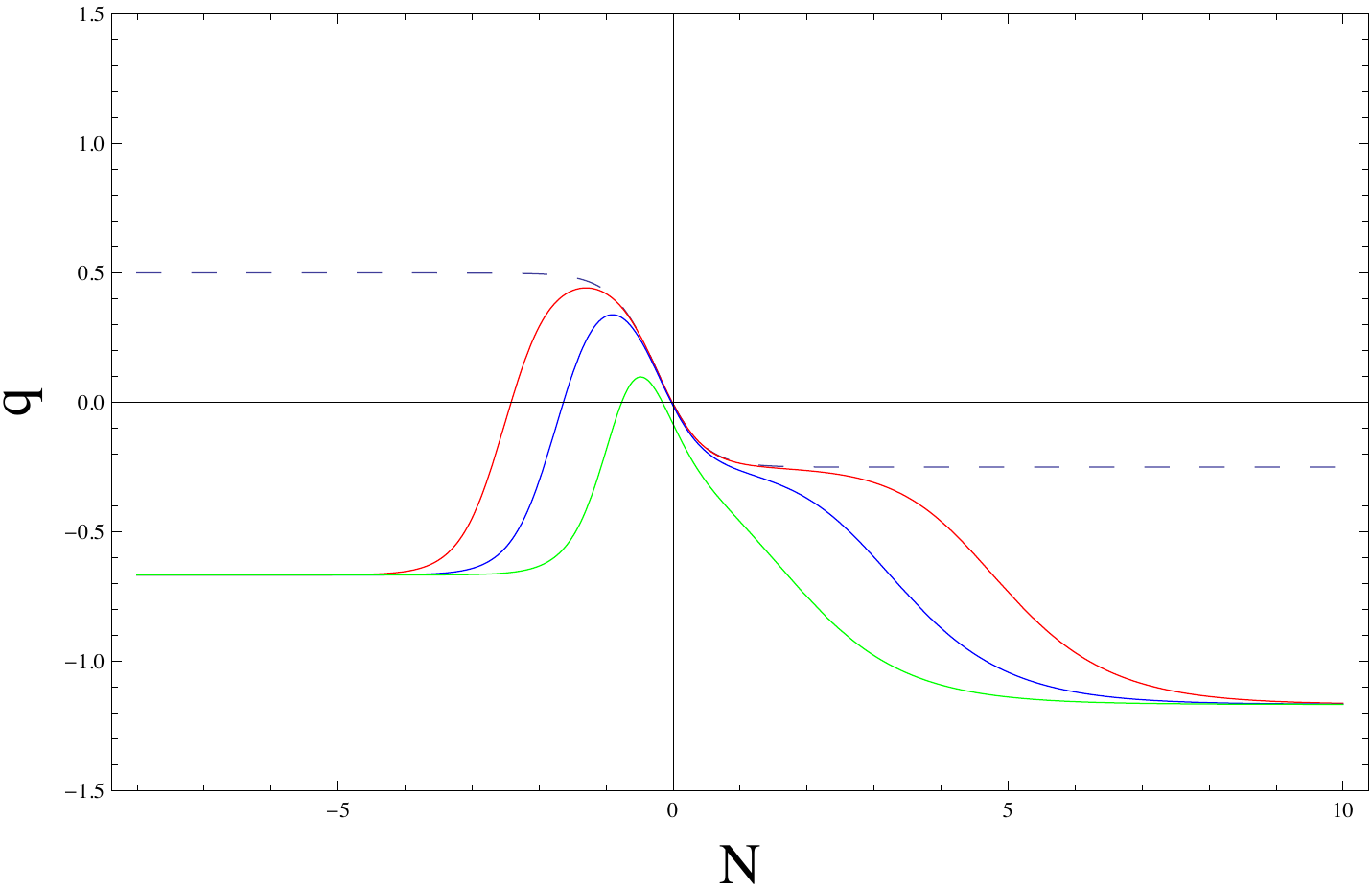} 
\end{tabular}
\caption{Numerical solution for Eq. \eqref{q}, where we have chosen the following initial conditions: $x_2(0)\equiv\Omega_p=0.6$ in the four plots and $\bar{\Omega}_m=0$, $\bar{\Omega}_p=0$ (Dashed plot), $\bar{\Omega}_m=10^{-4}$, $\bar{\Omega}_p=4\times10^{-4}$ (Red plot), $\bar{\Omega}_m=10^{-3}$, $\bar{\Omega}_p=4\times10^{-3}$ (Blue plot), $\bar{\Omega}_m=10^{-2}$, $\bar{\Omega}_p=4\times10^{-2}$ (Green plot). See the text for more details. } 
\label{figq}
\end{figure}

In addition, the deceleration parameter $q=-\ddot{a}/aH^2$ can be written in terms of Eqs. \eqref{eq:xi-yi_01} as:
\begin{equation}
q(N)=\frac{1}{2}\left(1-\frac{3}{2}x_2-\frac{7}{3}y_1-\frac{10}{3}y_2\right), \label{q}
\end{equation}
where we have used the Friedmann constraint to eliminate the $x_1$ variable. The corresponding plot can be seen in Fig. \ref{figq}, assuming the following initial conditions: $x_2(0)\equiv\Omega_p=0.6$ for the four plots and $y_1(0)\equiv\bar{\Omega}_m=0$, $y_2(0)\equiv\bar{\Omega}_p=0$ (Dashed plot), $\bar{\Omega}_m=10^{-4}$, $\bar{\Omega}_p=4\times10^{-4}$ (Red plot), $\bar{\Omega}_m=10^{-3}$, $\bar{\Omega}_p=4\times10^{-3}$ (Blue plot), $\bar{\Omega}_m=10^{-2}$, $\bar{\Omega}_p=4\times10^{-2}$ (Green plot) which are permitted small values according to the observations \cite{PlanckCollaboration2013}. Notice how brane terms, even if they are minimal, can cause an accelerated expansion process. Indeed, phantom dynamics in a non brane theory has a region where the Universe does not present an acceleration epoch, however, we can see that the more the brane effects are present the non-accelerated stages are less pronounced, which is clearly a contradiction to observations.

\begin{figure*}[htbp]
\centering
\begin{tabular}{cc}
\includegraphics[scale=0.24]{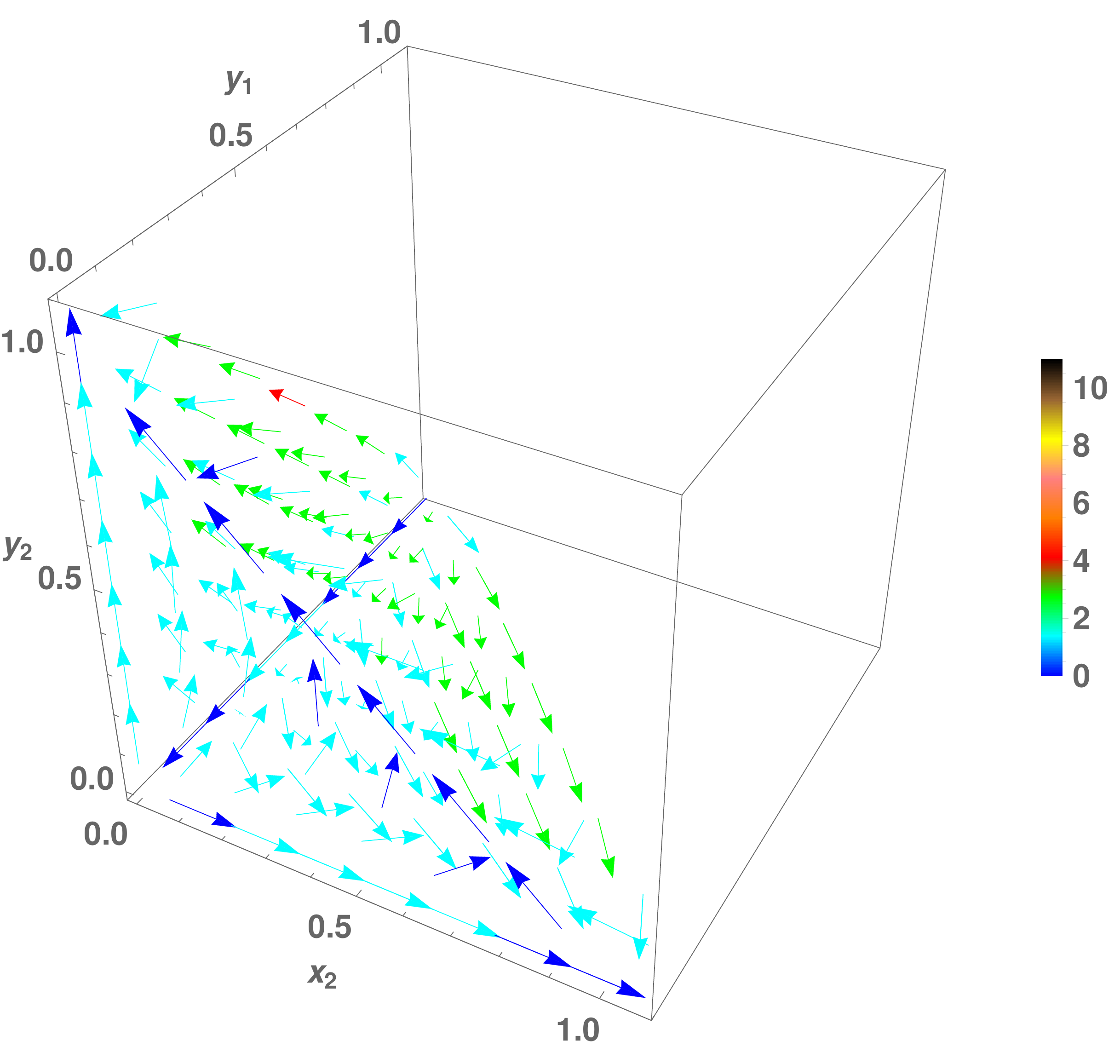} 
\includegraphics[scale=0.22]{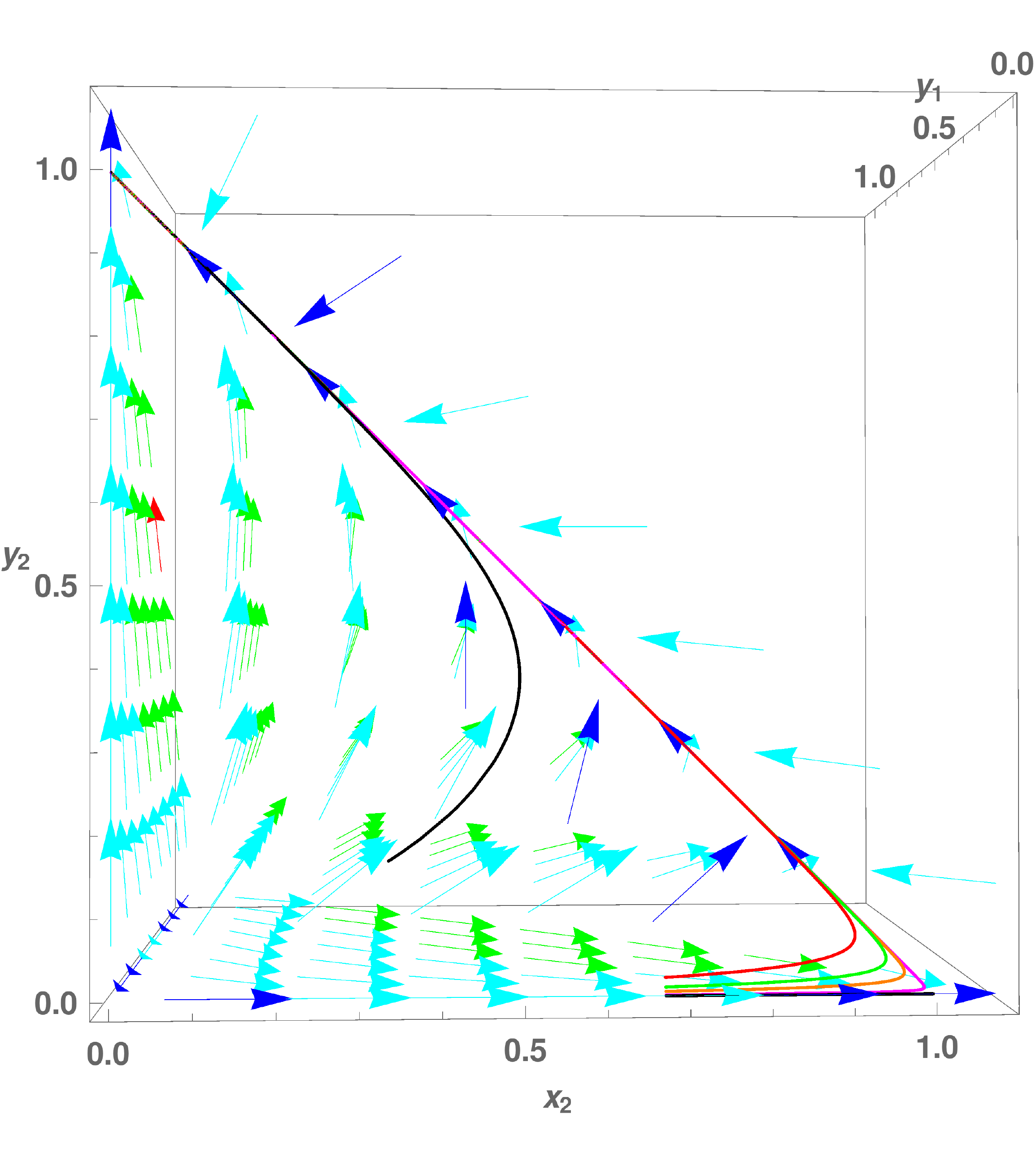} \\
\end{tabular}
\caption{Three dimensional vectorial field associated to the dynamical system of Eqs. \eqref{SDC}. The color bar indicates the magnitude of the vector, where we can se the presence of one repulser, two unstable points and one attractor related to the final stage of the Universe.} 
\label{fig3}
\end{figure*}

Another complementary analysis is shown in Figs. \eqref{fig3} where we present a vectorial dynamical analysis, showing only the region of interest, ergo, the region given by the Friedmann constriction. We start the analysis finding the equilibrium points and eigenvalues associated with Eq. \eqref{SDC}, defining the critical points as $(dx_i,y_i/dN)_{x_0}=0$. In this case as can be seen in Tab. \ref{tab}, the critical points are associated with matter domination, phantom DE domination, matter coupled with branes domination and phantom coupled with branes domination respectively.

In addition, we define the vector ${\bf x}=(x_2,y_1,y_2)$ and consider a linear perturbation of the form \eqref{bif0} and the Jacobian matrix $\mathcal{J}_{s_i}$ associated with the linearized system. The table \ref{tab} shows the eigenvalues and eigenvectors associated with the critical points of $\mathcal{J}_{s_i}$ Then, as seen in the previous section, the critical points can be classified according to the eigenvalues of the Jacobian of the linearized vector field at a specific point and also for the values below. Thus $(0,0,1)$ is an attractor and $(0,1,0)$ is a source since the eigenvalues associated with these points are all positive, while the origin (0,0,0) and $(1,0,0)$ are saddle points of the non-linear system since their eigenvalues have opposite signs. 

The region of interest is formed by families of solutions or dynamical fluxes, providing a qualitative description of the evolution of the system as a whole. The dynamics of a particular solution is governed by the initial conditions, $x_2(0)\equiv\Omega_{0p}$, $y_1(0)\equiv\bar{\Omega}_{0m}$, $y_2(0)\equiv\bar{\Omega}_{0p}$, which are called solution curves. In Figs. \ref{fig3} it is shown the vector field and some numerical solutions (solid lines) for different initial conditions, all of them satisfying the Friedmann constriction.

\begin{table}[htp]
\caption{Table of eigenvalues and eigenvectors associated to each critical point of the Jacobian matrix, for k=1.\label{tab}  }
\begin{ruledtabular}
\begin{tabular}{c c c c c c }
Critical points & Eigenvalues & Eigenvectors  \\
$s_0$=(0,0,0) & \{6,9/2,-3\} & (0,0,1), (1,0,0), (0,1,0)  \\
$s_1$=(1,0,0) & \{-15/2,-9/2,3/2\} & (-1,1,0), (1,0,0), (-1,0,1)  \\
$s_2$=(0,1,0) & \{9,15/2,3\} & (0,-1,1), (-1,1,0), (0,1,0) \\
$s_3$=(0,0,1) & \{-9,-6,-3/2\} & (0,-1,1), (0,0,1), (-1,0,1)
\end{tabular}
\end{ruledtabular}
\end{table}

\section{Phantom Dark Energy: A refined analysis} \label{PF}

In the previous sections we only considered the phantom field under the constriction $\omega_p<-1$, however we now propose a deeper analysis through the explicit form of the phantom field.

The coupling with gravity is given by the action \cite{Copeland:2006wr}
\begin{equation}
S[g,\phi]=\int d^4x\sqrt{-g}\left[\frac{1}{2}(\nabla\phi)^2-V(\phi)\right],
\end{equation}
with an opposite sign in the kinetic term, where $\phi$ is the phantom scalar field. Therefore, density and pressure are written as $\rho=-\dot{\phi}^2/2+V(\phi)$ and $p=-\dot{\phi}^2-V(\phi)$.

Furthermore, as we previously mentioned, it is possible to avoid the Big Rip singularity if the potential:
\begin{equation}
V(\phi)=V_0[\cosh(\sqrt{G_N}\beta\phi)]^{-1},
\end{equation}
has a maximum, where $\beta$ is a constant \cite{Singh}.

\begin{figure}[htbp]
\centering
\begin{tabular}{cc}
\includegraphics[scale=0.47]{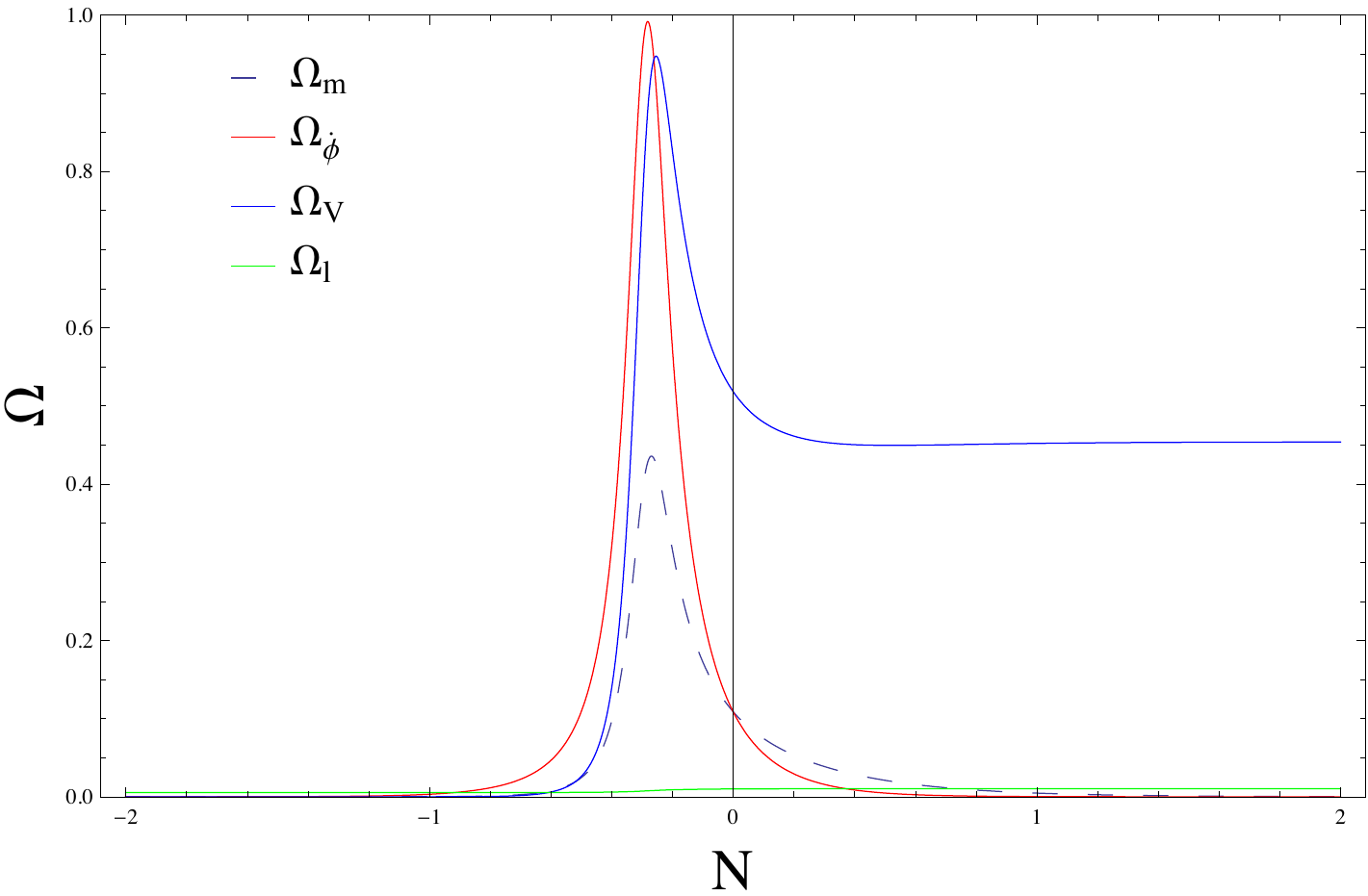} 
\end{tabular}
\caption{Numerical solution of Eqs. \eqref{sys2}, under the initial conditions: $\Omega_m=0.33$, $\Omega_{\dot{\phi}}=0.33$, $\Omega_{V}=0.72$ and $\Omega_{l}=10^{-1}$. Notice that phantom field dominates for large e-foldings.} 
\label{figPhant}
\end{figure}

Thus, the Friedmann equation can be written as:
\begin{widetext}
\begin{equation}
H^2=\frac{8\pi G_N}{3}\left\lbrace\rho_m\left(1+\frac{\rho_m}{2\lambda}\right)-\frac{1}{2}\dot{\phi}^2\left(1-\frac{\dot{\phi}^2}{4\lambda}\right)+\frac{V_0}{\cosh(\sqrt{G_N}\beta\phi)}\left(1+\frac{1}{2\lambda}\left[\frac{V_0}{\cosh(\sqrt{G_N}\beta\phi)}-\dot{\phi}^2\right]\right)\right\rbrace, \label{FriedPhan}
\end{equation}
\end{widetext}
along with the following equations
\begin{eqnarray}
\ddot{\phi}+3H\dot{\phi}+\partial_{\phi}V=0, \;\;\; \dot{\rho}_m+3H\rho_m=0.
\end{eqnarray}
Now, defining the appropriate dimensionless equations:
\begin{subequations} \label{dim2}
\begin{eqnarray}
&&x^2\equiv\frac{8\pi G_N\rho_m}{3H^2}, \;\; y^2\equiv\frac{4\pi G_N\dot{\phi}^2}{3H^2}, \;\; k^2\equiv\frac{3H^2}{16\pi G_N\lambda},\\
&&u^2\equiv\frac{8\pi G_NV_0}{3H^2\cosh(\sqrt{G_N}\beta\phi)}, \;\; \\
&&l^2\equiv\sqrt{\frac{3}{4\pi}}\beta\tanh(\sqrt{G_N}\beta\phi),
\end{eqnarray}
\end{subequations}
it is possible to reduce the modified Friedmann Eq. \eqref{FriedPhan} to:
\begin{equation}
1=x^2+(u^2-y^2)+k^2[x^4+(u^2-y^2)^2],
\end{equation}
recovering the traditional Friedmann equation when $k\to0$. Then, the dynamical system can be written as:
\begin{subequations} \label{sys2}
\begin{eqnarray} 
&&x^{\prime}=-\left(\frac{3}{2}+\frac{H^{\prime}}{H}\right)x, \\
&&y^{\prime}=-\left(3+\frac{H^{\prime}}{H}\right)y+\frac{1}{2}u^2l^2, \\
&&u^{\prime}=-\left(\frac{1}{2}l^2y+\frac{H^{\prime}}{H}\right)u,
\end{eqnarray}
\end{subequations}
altogether with
\begin{equation}
l=\Omega_l\left[\sigma\tanh\left(2\sigma\int y^2dN\right)\right]^{1/2},
\end{equation}
where $\sigma\equiv\sqrt{3/4\pi}\beta$ is another free parameter, which must be assigned in order to solve the previous equations.  Notice that we also made use of the Friedmann constriction. Additionally we have:
\begin{eqnarray}
\frac{H^{\prime}}{H}&=&-\frac{3}{2}x^2-\frac{1}{2}u^2l^2y-\frac{1-x^2-(u^2-y^2)}{x^4+(u^2-y^2)^2}(6y^4-2u^2l^2y^3\nonumber\\&&-6u^2y^2+\frac{3}{2}u^4l^2y).
\end{eqnarray}
Therefore the critical points for the system (\ref{sys2}) are $(\pm1,0,0),(0,\frac{3\pm\sqrt{9-l^4}}{l^2 },1),(0,\frac{3\pm\sqrt{9-l^4}}{l^2 },-1)$ and it is possible to plot the dynamical system as we show in Fig. \ref{figPhant} with the initial conditions: $\Omega_m=0.33$, $\Omega_{\dot{\phi}}=0.33$, $\Omega_{V}=0.72$ and $\Omega_{l}=10^{-1}$, obtaining the expected behavior for a matter domination at earlier times and a posterior domination of the phantom DE, mainly in the potential of the field (see Fig. \ref{figPhant}); immediately we recognize a state where $V\gg\dot{\phi}^2$ for large values of $N$, producing an accelerated state.

\begin{figure}[htbp]
\centering
\begin{tabular}{cc}
\includegraphics[scale=0.5]{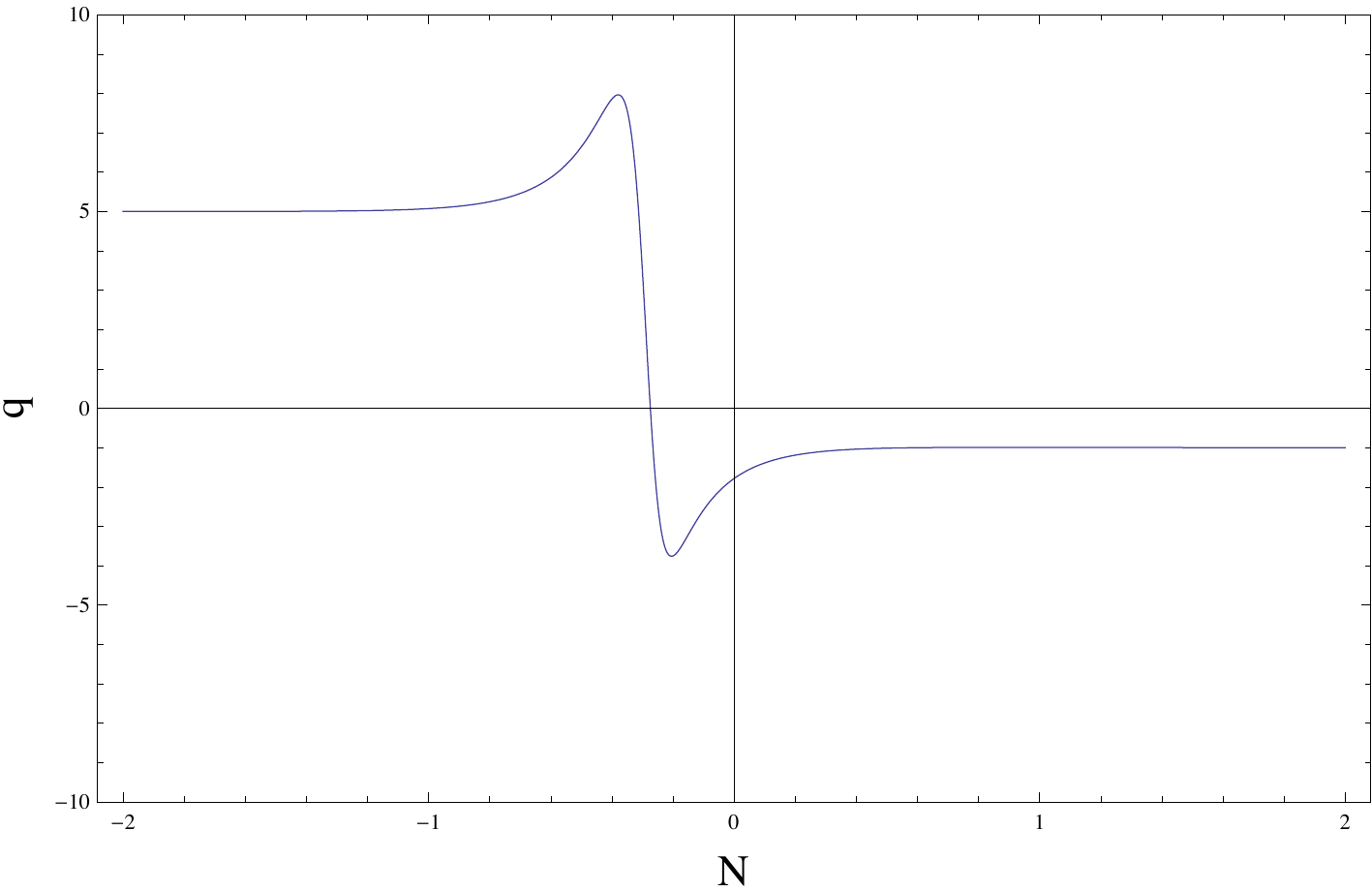} 
\end{tabular}
\caption{Numerical solution of Eq. \eqref{qPhanto}, under the initial conditions: $\Omega_m=0.33$, $\Omega_{\dot{\phi}}=0.33$, $\Omega_{V}=0.72$ and $\Omega_{l}=10^{-1}$. It is possible to observe a sudden acceleration when $N\sim-0.4$.} 
\label{qPhant}
\end{figure}

Another conclusive study can be performed through the deceleration parameter $q(N)$ which can be written in terms of the dimensionless variables \eqref{dim2} as:
\begin{eqnarray}
q(N)&=&\frac{1}{2}x^2-2y^2-u^2+\frac{1-x^2-(u^2-y^2)}{x^4+(u^2-y^2)^2}\times\nonumber\\&&(2x^4+5y^4-4y^2u^2-u^4), \label{qPhanto}
\end{eqnarray}
where its behavior is shown in Fig. \ref{qPhant}. From here, it is possible to observe a transition phase between an unaccelerated and accelerated state at $N\sim-0.4$ as would be expected in the traditional Universe behavior. However our results show a sudden phase transition in a short region of $N$, remaining stable for the value of $q\simeq-1$ and a Universe in continuous state of acceleration.

\begin{figure}[htbp]
\centering
\begin{tabular}{cc}
\includegraphics[scale=0.47]{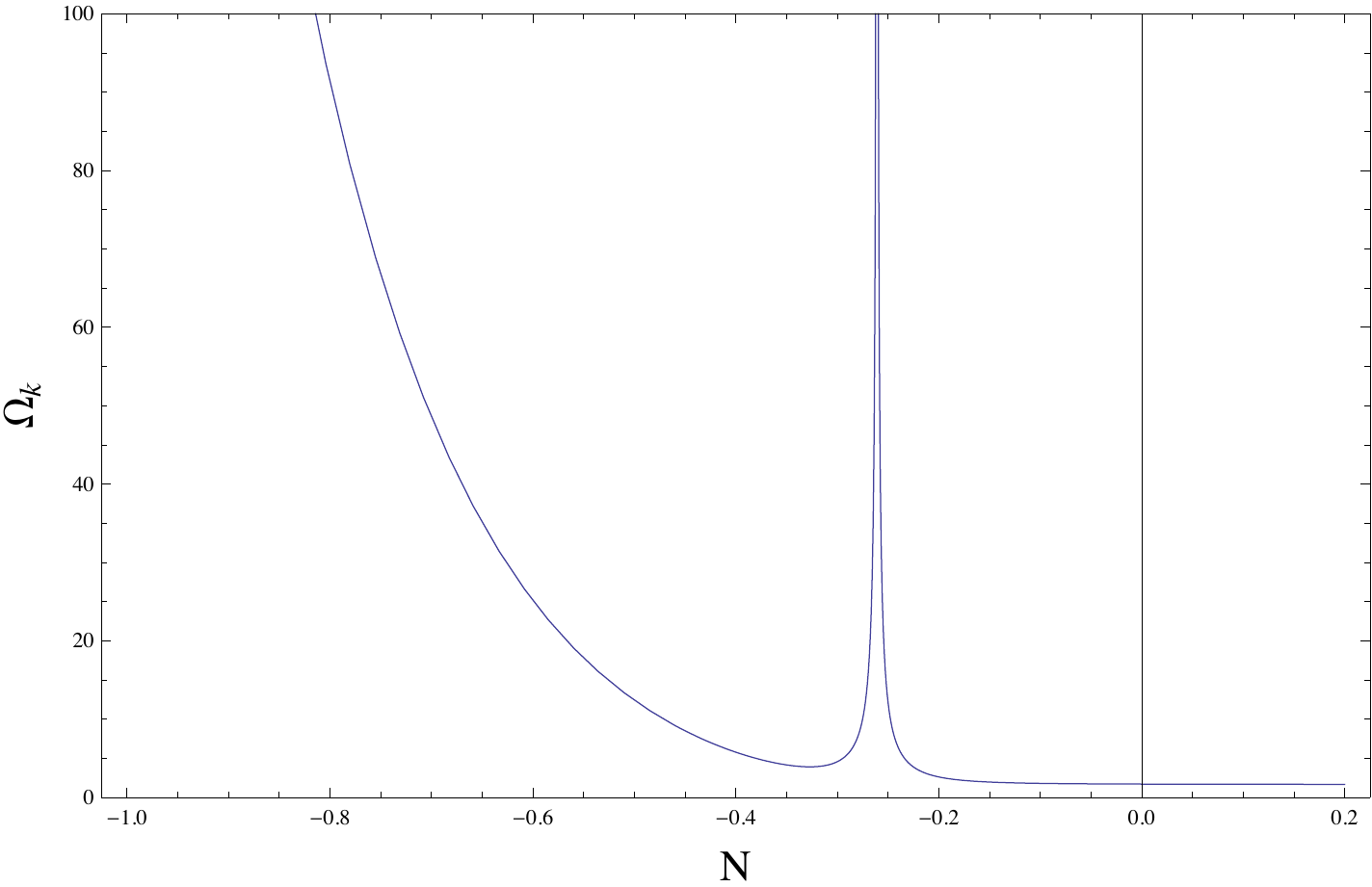} 
\end{tabular}
\caption{Numerical solution of Eq. \eqref{qPhanto}, under the initial conditions: $\Omega_m=0.33$, $\Omega_{\dot{\phi}}=0.33$, $\Omega_{V}=0.72$ and $\Omega_{l}=10^{-1}$. It is possible to observe an abrupt peak between $N\sim-0.4$ and $N\sim-0.2$, associated to the transition between of an unaccelerated to an accelerated Universe. See the text for details.} 
\label{k}
\end{figure}

Additionally, some extra information comes from the $k$ parameter which has a dynamical equation $k^{\prime}=kH^{\prime}/H$, related with the brane tension. The numerical solution can be observed in Fig. \ref{k}, where we see a domination of the brane tension component in the earlier times of the Universe evolution, along with an abrupt peak related to the transition between an unaccelerated and accelerated Universe, and finally a subdominant epoch at later times. 
This evolution is always constrained with the brane tension which is bounded by current observations \cite{MaartensR,Garcia-Aspeitia:2016kak}. 

\begin{figure*}[htbp]
\centering
\begin{tabular}{cc}
\includegraphics[scale=0.5]{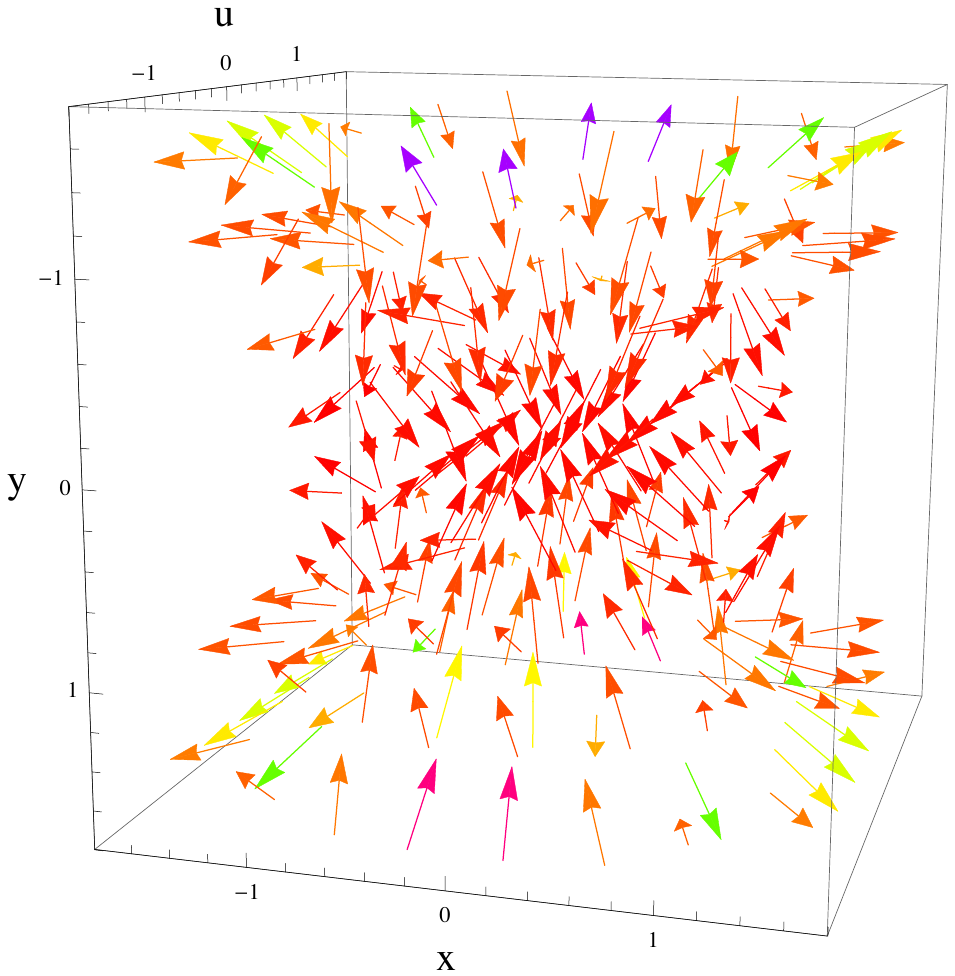} 
\includegraphics[scale=0.5]{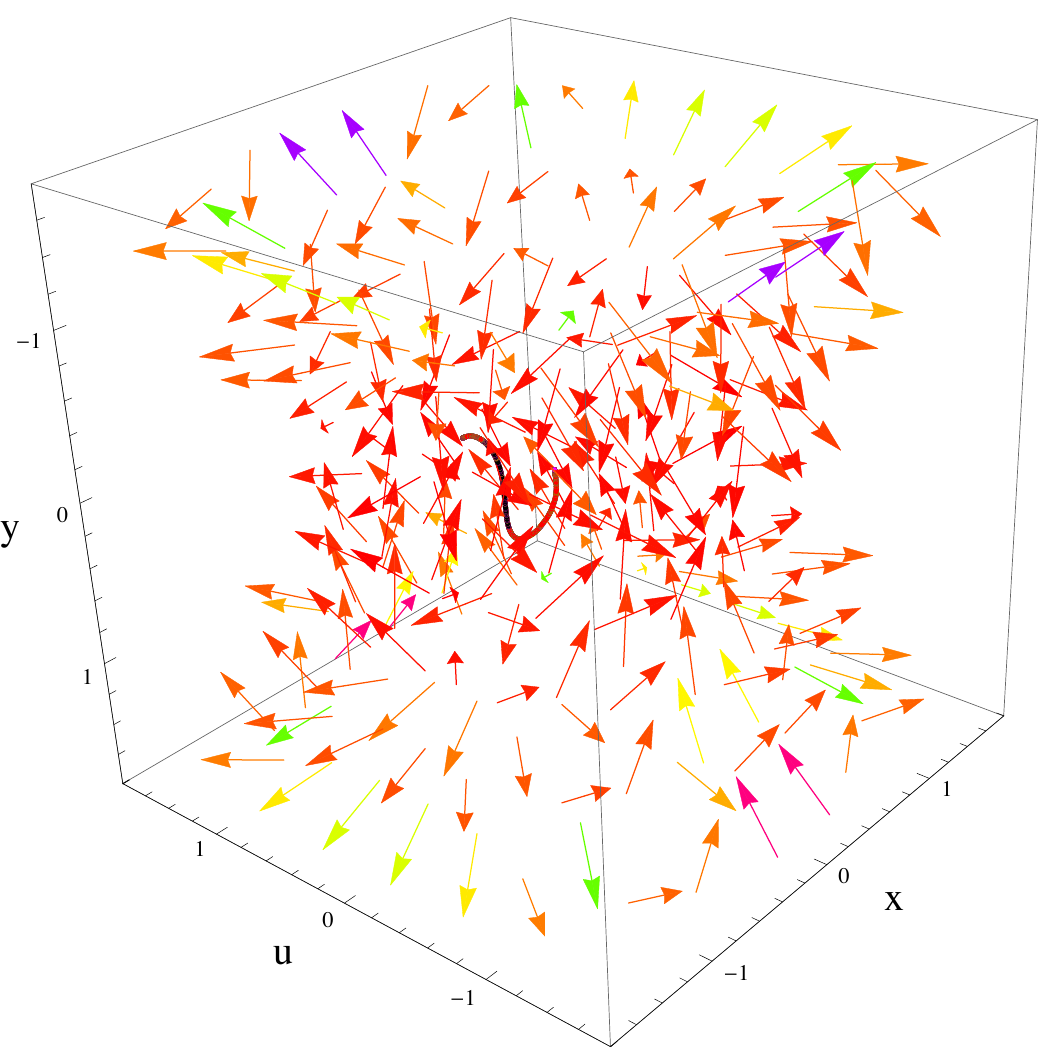} \\
\end{tabular}
\caption{Three dimensional vectorial field associated to the dynamical system of Eqs. \eqref{sys2}; fixing the parameters $l=10^{-1}$ and $k=10^{-1}$ by TT experiments.} 
\label{figX}
\end{figure*}

Finally, we explore the vector field of the system \eqref{sys2}, fixing the variables $l=k=10^{-1}$, which is shown in Fig. \ref{figX}; our results present two repulsers (the baryonic matter and the kinetic part of the phantom) and one attractor associated to the phantom DE potential.

\section{Conclusions and Discussion} \label{CD}

The results presented in Figs. \ref{fig1} show that branes generates a premature Big Rip singularity while the brane tension is accentuated. We emphasize, that this important result can not be obtained in the traditional cosmological analysis with phantom dark energy.
In this vein, the analysis developed in this paper, unmasks the dominant components in the beginning and in the end of the Universe (see Figs. \ref{fig2}), showing that the density parameters in function of the brane tension will dominates in the future. The previous results are corroborates by Figs. \ref{fig3} where it is shown the repulsers associated with the components in the beginning of the Universe and the attractors related with the presence of extra dimensions, in future epochs (see also Table \ref{tab}).

Moreover, some important extra information can be obtained from the deceleration parameter $q$, showing the differences with the standard cosmological model ($\Lambda$CDM). Here it is possible to observe an ever-accelerating Universe, as the presence of the extra dimension increases. These results from the deceleration parameter agree with those shown in previous figures. In addition, the TT experiments or others, could restrict the dynamics of the presence of extra dimensions, mimicking to a large extent, the standard cosmological model. 

As a complement, we develop an analysis, when the form of the phantom DE is explicitly written as a scalar field. Indeed, we assume the same potential used in Ref. \cite{Singh}, with the aim of avoiding a Big Rip singularity, but now in the brane-world context.
Thus, this scenario generates a matter dominant era and a posterior domination of the phantom field through the variable $\Omega_V$, which depends on the SF potential, implying an accelerated Universe, for values $\omega\sim-1$. In addition to this, the deceleration parameter (see Fig. \ref{qPhant} ) also give us information about the Universe passing from a non accelerated to an accelerated state (which is an expected result), having an abrupt change of phase (decelerated$\to$accelerated) between $N=-0.8$ and $N=0.2$; coinciding with the region where the brane tension presents an anomalous behavior (see Fig. \ref{k}). It is also possible to discuss that in Fig. \ref{k} the density parameter related with brane tension presents the expected behavior with a dominant brane tension in early epochs and subdominant brane tension in late epochs.
Finally, the vector field presented in Fig. \ref{figX} corroborates our results, presenting the expected attractor related to phantom dark energy potential and a repulser in the early times of the Universe evolution.

As a final comments, phantom field in brane-world scenario generates a premature Big Rip by the presence of brane tension, which can be stopped with a the presence of a potential with a local maximum. This last analysis show us a more adequate and congruent behavior as expected by observations, except for the transition region. Further analysis with observations will help us to constrict the brane tension, however this is work that will be presented elsewhere.

\begin{acknowledgements}
We would like to thank the referees for thoughtful comments which helped to improve the manuscript. ROA-C and JAA-M acknowledge support from PhD CONACYT fellowship, MAG-A acknowledge support from SNI-M\'exico and CONACyT research fellow. We acknowledge as well the invaluable support of Cesar Martinez, due to his careful reading of the first draft and suggestions. We also want to thank Instituto Avanzado de Cosmolog\'ia (IAC) collaborations. 
\end{acknowledgements}

\bibliography{librero1}

\end{document}